\documentclass[12pt]{revtex4}

\usepackage{graphicx}
\usepackage{dcolumn}
\usepackage{bm}

\begin{document}

\title{House Price Distributions of Taiwan: A Preliminary Study}

\author{Chung-I Chou$^1$ and S. P. Li$^2$}

\affiliation{$^1$Department of Physics, Chinese Culture University, Taipei, Taiwan 111, R.O.C.}
\affiliation{$^2$Institute of Physics, Academia Sinica, Taipei, Taiwan 115, R.O.C.}

\begin{abstract}
The house price distributions of Taiwan are analyzed.  The tail of the cumulative distribution
function (CDF) follows an approximate power law with an exponent equals to -2.4 while the 
distribution of the house price per unit area displays a lognormal distribution.  Implications 
of the results are also discussed.  
\end{abstract}

\maketitle

Many complex systems exhibit heavy-tailed distributions in observables that characterize 
the systems.  Among them are natural hazards such as earthquakes, landslides, wildfires \cite{Malamud}
or the ranking of words used in literature \cite{Zipf}.  In a recent work \cite{Ohnishi}, it was realized
that the tail of the house price distribution in the Greater Tokyo Area differs significantly from
a lognormal distribution as would have been assumed by many people, and in fact 
display a power law behavior.  One would want to ask if this kind of behavior is universal
or just a special case related to the Greater Tokyo Area.  Indeed, a better
understanding of the evolution of real estate prices such as the house price distribution 
might help the government to have a better control over the abnormal price fluctuations 
and thus to avoid economic downturns before they really happen.  In this 
preliminary study, we report the distribution of house price in Taiwan in 2010.  
In order for Taiwanese citizens to gather better information about the house price 
in various areas in Taiwan, its government started in 2010 to regularly release the data of the price
of house that are recently sold.  The first set of released data is the price of house 
sold in the first quarter of 2010 in Taiwan \cite{DataBank1}.  In this set of data, there are a total of 
7,374 units (including houses, apartments, factories, etc) sold during this period.  
Of these seven thousand plus units, only 6,696 of them are for residential use so we 
will only include these six thousand plus units in the following analysis.  

Figure 1 is the log-log plot of the cumulative distribution function (CDF) of the 
residential house price in Taiwan in the first quarter of 2010.  This CDF includes both
the residential house price in urban as well as suburban and rural areas.  The x-axis
is the total price for the house sold and the unit here is NTD (New Taiwanese Dollar, 
1 USD $\approx$ 32 NTD) while the y-axis shows the number of cases sold.  
One can see that the tail of the cumulative distribution function 
significantly differs from an exponential or lognormal distribution.  In fact, the tail
can be approximated by a power law distribution with an exponent of about -2.4.  This 
result is very similar to that of the CDF obtained in \cite{Ohnishi}.  

\begin{figure}
\includegraphics{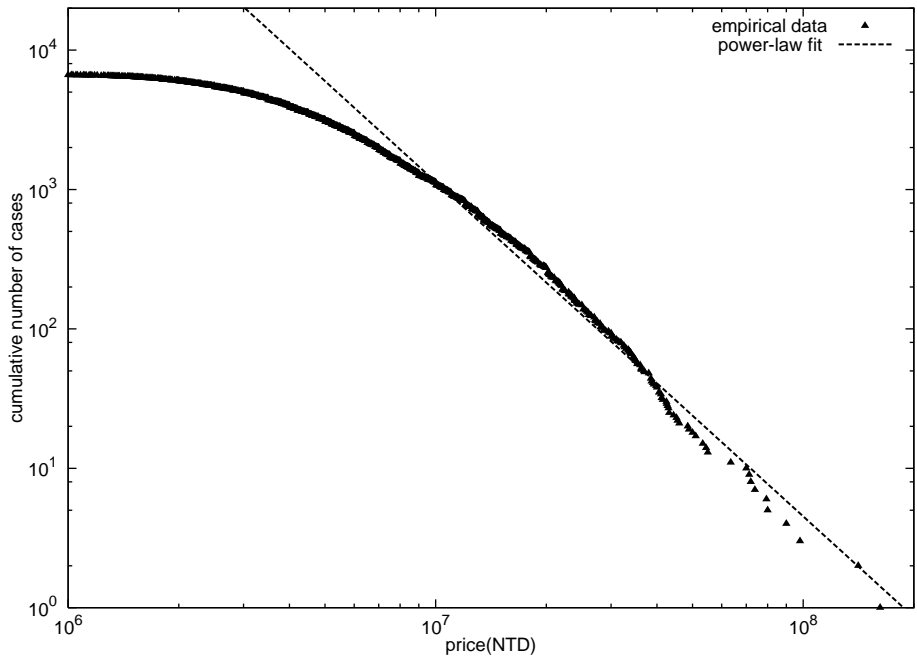}
\caption{\label{fig:1} The cumulative distribution function (CDF) of the 
residential house price in Taiwan in the first quarter of 2010. The tail exhibits an 
approximate power law behavior with an exponent of -2.4.  }
\end{figure}

Figure 2 is the house price per unit area in Taiwan in the same period.  The
data points in solid triangles show the distribution function for the 
houses sold throughout Taiwan during this period.  The x-axis is the price per unit area 
(NTD per meter squared) and the y-axis shows the number of cases sold.  At first glance, the distribution 
seems somewhat strange since there is a bump on the right hand side of the peak of the 
distribution. 

\begin{figure}
\includegraphics{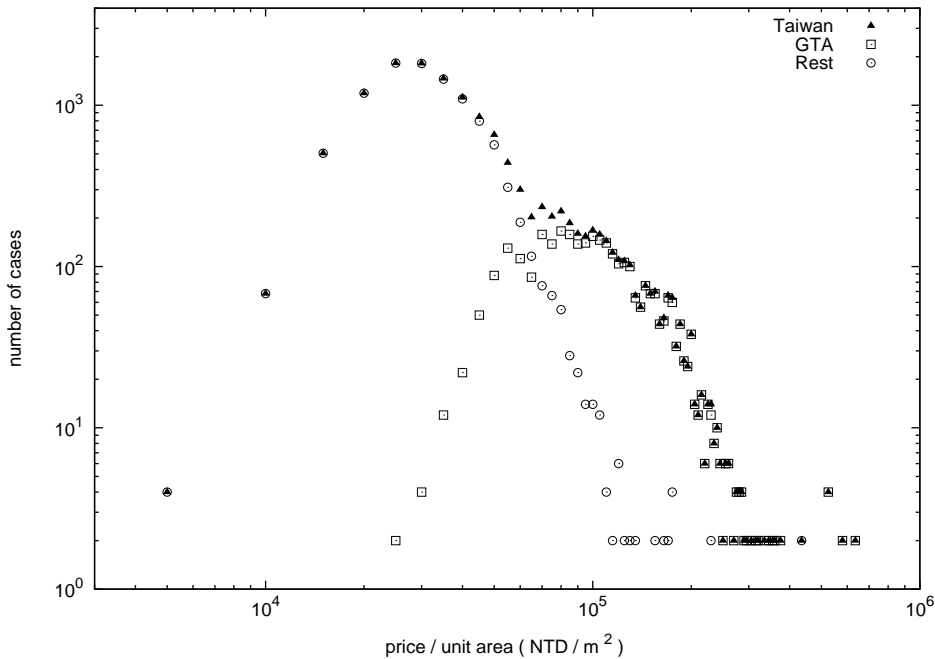}
\caption{\label{fig:2} The house price per unit area (NTD/m$^2$) in Taiwan in 2010.  The data points
in solid triangles are for the whole Taiwan area.  The data points in squares and circles 
are for the Greater Taipei Area (GTA) and for the rest of Taiwan respectively.  }
\end{figure}

Very recently, the Taiwanese government has announced a new 
restriction on home mortgages in Taiwan.  For the designated Greater Taipei Area, the same
person gets much stricter restrictions on home mortgage for the second house owned under 
his/her name.  This is because 
the house price in certain districts in the Greater Taipei Area has almost doubled
in the last two years.  To have a better understanding
of the distribution in Figure 2, we therefore divide the houses sold into two groups,
the designated Greater Taipei Area identified by the Taiwanese government and the
rest of Taiwan.  The PDF of the house price per unit area of these two groups are also
included in Figure 2, with the squares for the designated Greater Taipei Area and the
circles for the rest of Taiwan.  It is surprising to find out that they both follow
lognormal distributions though the positions of the peaks are different.  In fact, the
two lognormal distributions can be put into a single lognormal distribution if one makes
a shift of the peaks of the two distributions.  To do so, we multiply the house price 
of each house sold in Taiwan (other than in the designated Greater Taipei Area) by the same 
constant (ratio), in this case, 
about 3.3 which effectively means to translate the lognormal distribution to the right.  
One can now see that the two lognormal distributions collapse into a single lognormal(
$	\frac{ 1 }{x\sigma\sqrt{2\pi}} \exp(  \frac{-(ln(x)-\mu)^2}{2 \sigma^2 } ) $, with $\mu= 11.5$ and $\sigma=0.4$), as shown in 
Figure 3.  Notice that the y-axis here is normalized to represent the probability 
density function (PDF) of the distribution.  

\begin{figure}
\includegraphics{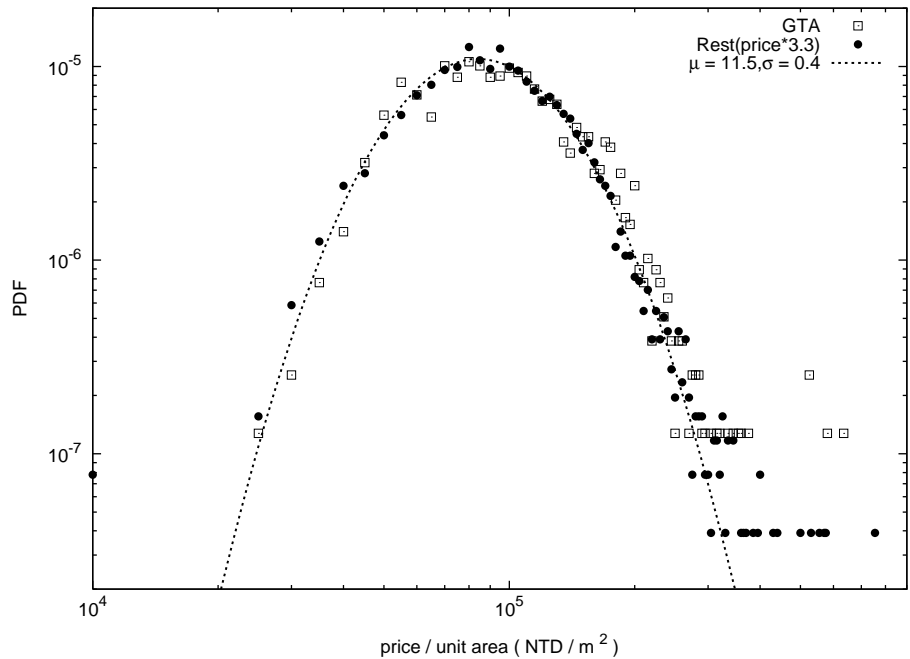}
\caption{\label{fig:3} Rescaled plot of Figure 2.  The lognormal distribution of the 
house price/unit area for houses sold in the first quarter of the rest of Taiwan area 
is rescaled by multiplying by a factor of 3.3.  The two lognormal distributions in 
Figure 2 now collapse into a single lognormal distribution.  }
\end{figure}

Incidentally, the released data from the government \cite{DataBank2} indicate that the average
income of a family in the Greater Taipei Area in 2009 was about 1.5 times that of the 
rest of Taiwan.  This means that houses in the Greater Taipei Area are indeed overpriced
by a factor of about 3.3/1.5 = 2.2!  A policy to correct this huge difference between 
the Greater Taipei Area and the rest of Taiwan by the
government is therefore urgent.  

\vskip 0.5cm

\noindent{Summary}--In this preliminary study of the house price distributions in Taiwan, we found that 
the tail of the cumulative distribution function (CDF) of the residential house price 
in Taiwan can be approximated by a power law distribution with an exponent of about -2.4.  
This is similar to the result obtained in a recent work [3] on the house price in the Greater Tokyo
Area.  On the other hand, the average house price per unit area in different regions 
of Taiwan follow lognormal distributions.  We have divided the houses sold into two 
groups--the designated Greater Taipei Area and the rest of Taiwan.  Adjusted by a ratio 
of 3.3, the two lognormal distributions indeed collapse into a single lognormal distribution.  
From this, we conclude that the houses sold in the Greater Taipei Area are overpriced by 
a factor of 2.2.  This will be indicated by the difference in the Misery 
Index between urban and suburban areas and will thus be a useful reference for 
policymakers.

\end{document}